\documentclass[12pt]{article}
\usepackage{amsmath, amsthm, amssymb,amsfonts}

\usepackage{hyperref}
\usepackage[nosort]{cite}
\def\be{\begin{equation}}
\def\ee{\end{equation}}
\def\O{{\mathcal O}}

\begin{document}
\begin{titlepage}
\begin{center}
{\bf\Large{\vbox{\centerline{A simple analysis of the mixed-state information metric} \vskip .5cm
 \centerline{in AdS$_3$/CFT$_2$}}}}
\vskip
0.5cm
{Shan-Quan Lan$^{a}$\footnote{shanquanlan@mail.bnu.edu.cn}, Gu-Qiang Li$^{a}$\footnote{zsgqli@hotmail.com}, Jie-Xiong Mo$^{a}$\footnote{mojiexiong@gmail.com}, Xiao-Bao Xu$^{a}$\footnote{xuxb@lingnan.edu.cn}} \vskip 0.05in
{\it ${}^a$ Institute of Theoretical Physics,\\ Lingnan Normal University, Zhanjiang, 524048 Guangdong, China }
\end{center}

\vskip 0.5in
\baselineskip 16pt
\abstract{We compute the quantum information metrics of a thermal CFT on $\mathbb R^{1,1}$ perturbed by the scalar primary operators of conformal dimension $\Delta=3,4,5,6$. In particular, we assume that the Hamiltonian of the mixed state commutes with each other and the temperature is fixed. Under these conditions, the evaluation is analogous to the pure state case. We also apply the method of [{arXiv:1607.06519}] to calculate the mixed state information metric for the scalar primary operator with conformal dimension $\Delta=4$ holographically. We find an exact agreement between the two results in our approach.}
\end{titlepage}

\section{Introduction}
Quantum information\cite{2010qcqi} has recently driven intense activity in understanding the AdS/CFT correspondence\cite{Maldacena:1997re,Gubser:1998bc,Witten:1998qj}. This relation between quantum information and AdS/CFT can be traced back to the remarkable proposal for a holographic expression of entanglement entropy in the boundary theory\cite{Ryu:2006bv}, also known as the Ryu-Takayanagi formula. An important application of this formula is that the evolution of time-dependent entanglement entropy in the CFT can be described by the holographic setups, Hartman and Maldacena\cite{Hartman:2013qma} found that the linear growth of the entanglement entropy is the same as the linear growth of the nice slices along the spacelike $t$-direction in the black hole interior. Another considerable progress in connecting quantum information to gravitational dynamics has been made by Susskind, who point out that the length of an Einstein-Rosen bridge corresponds to the computational complexity of a state in the CFT\cite{Susskind:2014rva}.

More recently, quantum information metric and its gravity dual was studied in\cite{MIyaji:2015mia}, which reveals that a quantum information metric in the CFT is closely related to the volume of a maximal time slice in the AdS bulk. This correspondence and holographic complexity\cite{Stanford:2014jda} have a certain feature in common, specifically it is shown that a quantum information quantity in the boundary field theory can be dual to the volume of a maximal co-dimension $one$ hypersurface in the AdS background. Therefore this new correspondence should be useful for the programme of bulk reconstruction\cite{Harlow:2018fse}, for example\cite{Caputa:2017yrh}.

The quantum information metric\cite{Braunstein:1994zz,Spehner:2014jmp}, or fidelity susceptibility, measures the distance between two quantum states. It is clear that this quantity is vanishing if and only if the two states are identical. More precisely, let us consider a generic one-parameter family of states, $\sigma$ is an arbitrary density matrix for this family of states, which has a tunable parameter $\lambda$. Now we perturb $\sigma$ by the parameter $\lambda$, the fluctuation $\lambda\rightarrow\lambda+\delta\lambda$ for infinitesimal $\delta\lambda$ will lead to a new density matrix $\rho$. Then quantum information metric $G_{\lambda\lambda}$ is defined as
\be
F(\sigma,\rho)=1-G_{\lambda\lambda}\delta\lambda^2+\O(\delta\lambda^3)\label{def}
\ee
where $F(\sigma,\rho)$ is the fidelity of the states $\sigma$ and $\rho$, defined as
\be
F(\sigma,\rho)=\rm tr\sqrt{\sqrt{\sigma}\rho\sqrt{\sigma}} \label{fi}
\ee
For the pure states, the fidelity\eqref{fi} reduce to the inner product $|\langle\sigma|\rho\rangle|$. Moreover, the overlap $|\langle\sigma|\rho\rangle|$ in a Euclidean CFT on $R^d$ can be described by the path integral when the CFT is deformed by a scalar primary operator $\O$ whose coupling jump across the interface $\tau=0$, where $\tau$ is the Euclidean time coordinate\cite{MIyaji:2015mia}.  And if the perturbation $\O$ is a marginal operator, the bulk geometry dual to this interface field theory is termed $Janus~solution$\cite{Bak:2003jk}, which is obtained from the AdS Einstein-scalar theory. Evaluating the on-shell action of this theory, we can find a precise correspondence between the both sides\cite{Bak:2015jxd}. However this method is not feasible for the deformations induced by non-marginal primary operators, because the bulk geometries dual to these perturbations are not simple as the Janus solution. The difficulty was circumvented by Trivella\cite{Trivella:2016brw} based on an ingenious perturbative approach in the AdS bulk. Inspired by this work, quantum information metrics in other cases have been studied\cite{Bak:2017rpp,Chen:2018vkw}.

In this paper we investigate the mixed quantum information metric in two dimensional thermal CFT and its gravity dual for various non-marginal primary operators. To simplify our calculation, we only consider the deformation that the density matrices $\sigma$ and $\rho$ are commutative, meanwhile the temperature of the thermal CFT is fixed. Under these conditions, the fidelity \eqref{fi} also reduce to a path integral\cite{Bak:2015jxd}. Thus we can compute the mixed quantum information metric in this case, we will focus on the deformed CFT by the scalar primary operators with conformal dimension $\Delta=3,4,5,6$. Actually this is almost the problem in section 5 of \cite{Trivella:2016brw}, where the author obtained the quantum information metric
with the marginal deformation for a thermofield-double state in $d=2$. We will see that the mixed quantum information metric with the marginal deformation in our case can also be given by the formula $(60)$ in\cite{Trivella:2016brw}. However the mixed quantum information metrics with the deformations induced by the scalar primary operators of conformal dimension $\Delta=3,5$ have some elusive results. To our satisfaction, the mixed quantum information metrics with the deformations induced by the scalar primary operators of conformal dimension $\Delta=4,6$ have some regular results. Due to the complete agreement between the quantum information metric with the marginal deformation for a thermofield-double state in $d=2$ and its gravity dual\cite{Trivella:2016brw}, we think that there is still a precise correspondence for the mixed quantum information metric with the non-marginal deformation. With this motivation, we compute the mixed information metric for the scalar primary operator with conformal dimension $\Delta=4$ holographically by adopting the method of\cite{Trivella:2016brw}, and find that the result in gravity is the same as the one from the CFT.

The paper is organized as follows. In the next section we briefly review the quantum information metric, especially on the mixed information metric. In section \ref{QIMcft}, we calculate the mixed information metrics for the scalar primary operators with conformal dimension $\Delta=3,4,5,6$ in the CFT.
In section \ref{QIMgr}, we try to get the mixed information metric for the scalar primary operator with conformal dimension $\Delta=4$ in gravity following the method of \cite{Trivella:2016brw}. Finally, in section \ref{dis} we discuss the possible generalization of our consideration.

\section{Mixed-state information metric}
\label{QIM}
In this section, we briefly review some results concerning the mixed-state information metric. The quantum fidelity between two mixed states $(\rho_1,\rho_2)$ defined by
\be
F(\rho_1,\rho_2)=\rm tr\sqrt{\rho_1^{1\over 2}\rho_2\rho_1^{1\over 2}}
\ee
where $\rho_i$ is the density matrix of a thermal state given as
\be
\rho=\frac{1}{Z}\sum_n e^{-\beta E_n}|\Psi_n\rangle\langle\Psi_n|=\frac{1}{Z}e^{-\beta H}
\ee
In the present discussion, we consider that the thermal states $\rho_1, \rho_2$ belong to a family of the quantum states labeled by the parameter $\lambda$, in particular
\be
H_i=H(\lambda_i),\; \lambda_2=\lambda_1+\delta \lambda,\; \beta_2=\beta_1=\beta
\ee
where we have restricted that the thermal states $\rho_1, \rho_2$ are in the same temperature $1/\beta$.

Furthermore, we assume that the Hamiltonians $H_1,H_2$ are commutative, then
\be
F(\rho_1,\rho_2)=(Z_1 Z_2)^{-1/2}\,{\rm tr\,exp}\left[-\beta\left(H_1+H_2\right)/2\right] \label{tfi}
\ee
Before obtaining the path integral representation of the fidelity \eqref{tfi} in a thermal CFT, we first recall the quantum information metric for the ground state of a CFT on ${\rm R}^d$.

Let us start by considering the CFT with a Euclidean Lagrangian $\mathcal L_1$ on ${\rm R}^d$, whose Euclidean time and space coordinates are denoted by
$\tau$ and $x$. Next we deform the theory by a primary operator $\O$ of conformal dimension $\Delta$ at $\tau=0$. The new Lagrangian is
\be
\mathcal L_2=\mathcal L_1+\delta\lambda\,\O
\ee
where $\delta\lambda$ is a coupling constant. The overlap between the ground state of the undeformed theory $|\Psi_1\rangle$ and the ground state of the
deformed theory $|\Psi_2\rangle$ is
\be
\langle\Psi_2|\Psi_1\rangle=\frac{1}{(Z_1Z_2)^{1/2}}\int\mathcal D\phi\,{\rm exp}[-\int d^{d-1}x(\int_{-\infty}^0 d\tau\mathcal L_1+\int_0^{\infty}d\tau\mathcal L_2)]
\ee
Due to the sudden change in the action at $\tau=0$, this path integral is UV divergent, we need regularization and renormalization. We consider the regularization as follows
\be
|\Psi_2^\epsilon\rangle=\frac{e^{-\epsilon H_1}|\Psi_2\rangle}{(\langle\Psi_2|e^{-2\epsilon H_1}|\Psi_2\rangle)^{1\over 2}}
\ee
The regularized overlap can be written as
\be
\langle\Psi_2^\epsilon|\Psi_1\rangle=\frac{\langle{\rm exp}(-\int_\epsilon^\infty d\tau\int d^{d-1}x \,\delta \lambda \O)\rangle}{\langle{\rm exp}(-(\int_{-\infty}^{-\epsilon}+\int_\epsilon^\infty) d\tau\int d^{d-1}x \,\delta \lambda \O)\rangle^{1/2}} \label{rfi}
\ee
Expanding the regularized overlap \eqref{rfi} in small $\delta \lambda$, we can find that
\be
G_{\lambda\lambda}=\frac{1}{2}\int d^{d-1}x_1\int d^{d-1}x_2\int_{-\infty}^{-\epsilon}d\tau_1\int_\epsilon^\infty d\tau_2 \langle\O(\tau_1,x_1)\O(\tau_2,x_2)\rangle \label{purqim}
\ee
For the thermal CFT on $\rm R^{1,d-1}$, it can be formulated on $S^1_\beta\times \rm R^{d-1}$ by using the Imaginary-Time Formalism, where $\beta$ is the inverse temperature of the system. Similar to the setup in the Euclidean CFT on $\rm R^d$, we deform the original theory with the Lagrangian $\mathcal L_1$ at $\tau=0$ by the term $\delta\lambda\O$, and the original theory associated to the density matrix $\rho_1$ is defined on the interval $\mathcal I_1=[-\frac{\beta}{2},0]$, the deformed theory associated to the density matrix $\rho_2$ is defined on the interval $\mathcal I_2=[0,\frac{\beta}{2}]$.

Thus, the fidelity \eqref{tfi} can be rewritten by the path integral as
\be
F(\rho_1,\rho_2)=\frac{1}{(Z_1Z_2)^{1/2}}\int\mathcal D\phi\,{\rm exp}[-\int d^{d-1}x(\int_{-\frac{\beta}{2}}^0 d\tau \mathcal L_1+\int_0^{\frac{\beta}{2}}d\tau\mathcal L_2)]
\ee
Because of the periodicity of the thermal circle $S^1_\beta$, we need introducing the cut-off at both ends  of the interval $\mathcal I_2$, i.e.
$\mathcal I_2(\epsilon)=[\epsilon,\frac{\beta}{2}-\epsilon]$. Then the regularized density matrix of the deformed theory is described as
\be
\rho_2^\frac{1}{2}(\epsilon)=\frac{e^{-\epsilon H_1}e^{-(\frac{\beta}{2}-2\epsilon)H_2}e^{-\epsilon H_1}}{\sqrt{Z_2(\epsilon)}}
\ee
where $Z_2(\epsilon)={\rm tr}(e^{-\epsilon H_1}e^{-(\frac{\beta}{2}-2\epsilon)H_2}e^{-\epsilon H_1})^2$ \cite{Bak:2015jxd}.

Along the approach to driving the quantum information metric for the pure state \eqref{purqim}, the mixed-state information metric is given by
\be
G_{\lambda\lambda}=\frac{1}{2}\int_\epsilon^{\frac{\beta}{2}-\epsilon}d\tau_2 \int_{-\frac{\beta}{2}+\epsilon}^{-\epsilon} d\tau_1\int dx_2^{d-1}
\int dx_1^{d-1}\langle\hat{\O}(\tau_2,x_2)\hat{\O}(\tau_1,x_1)\rangle \label{mqim}
\ee
with $\hat{\O}(\tau,x)\equiv\O(\tau,x)-\langle\O(\tau,x)\rangle$. For a closer look at the mixed-state information metric \eqref{mqim}, we will focus on the two-dimensional CFT. If the primary operator $\O$ is not in the Virasoro vacuum module $\{1, T, \partial T,:T\,T:, \cdots\}$, we have \cite{Iliesiu:2018fao}
\be
\langle\O\rangle_{S^1_\beta\times \rm R}=0
\ee
In the following discussion, we will be interesting in those operators, therefore we get
\be
G_{\lambda\lambda}=\frac{1}{2}\int_\epsilon^{\frac{\beta}{2}-\epsilon}d\tau_2 \int_{-\frac{\beta}{2}+\epsilon}^{-\epsilon} d\tau_1\int dx_2^{d-1}
\int dx_1^{d-1}\langle\O(\tau_2,x_2)\O(\tau_1,x_1)\rangle \label{2pt}
\ee
In order to compare the $\rm CFT_2$ results with the gravitational one in $AdS_3$, we choose the two-point function as
\be
\langle\O(\tau_1,x_1)\O(\tau_2,x_2)\rangle=\mathcal C \frac{(\frac{\pi}{\beta})^{2\Delta}}{[\sinh^2(\frac{\pi(x_1-x_2)}{\beta})+ \sin^2(\frac{\pi(\tau_1-\tau_2)}{\beta})]^\Delta}
\ee
where $\mathcal C=\frac{4\eta l(\Delta-1)}{\pi}$, with $\eta=\frac{1}{16\pi G}$ denoting the $d+1$ dimensional Newton's constant G, and $l$ is the AdS radius, please see the eq.(3.2) in \cite{Bak:2017rpp}.

\section{Mixed-state information metric in CFT$_2$}
\label{QIMcft}
In this section we will perform the calculation of the mixed-state information metrics for the scalar primary operators $\O$ with conformal dimension
$\Delta=3,4,5,6$ according to Eq. \eqref{2pt}. The mixed-state information metric for the scalar primary operator $\O$ with conformal dimension $\Delta=2$
has been derived by Trivella, who also find it through AdS/CFT \cite{Trivella:2016brw}.

$\bullet\quad \Delta=3$
\begin{align}
G_{\lambda\lambda}=&\frac{V_{\mathbb R}}{2}\mathcal C(\frac{\pi}{\beta})^3\int_{-\frac{\pi}{\beta}\tilde{\tau}+\epsilon' }^{\frac{\pi}{2}+\frac{\pi}{\beta}\tilde{\tau}-\epsilon' }d\tau_1'\int_{\frac{\pi}{2}+\frac{\pi}{\beta}\tilde{\tau}+\epsilon'}^{\pi-\frac{\pi}{\beta}\tilde{\tau}-\epsilon' }d\tau_2'\int_{-\infty}^{+\infty}dx_1'\frac{1}{(\sinh^2(x_1'-x_2')+\sin^2(\tau_1'-\tau_2'))^3}\cr
=&\frac{V_{\mathbb R}}{2}\mathcal C(\frac{\pi}{\beta})^3\Big\{(-4)\int_{2\epsilon'}^{\frac{\pi}{2}-\frac{2\pi}{\beta}\tilde{\tau}}du\;(2u-4\epsilon')[\frac{3(2u-\pi)}{ \sin^5(2u)}-\frac{2(2u-\pi)}{\sin^3(2u)}
-\frac{3\cos(2u)}{\sin^4(2u)}]\cr
&-2(\pi-\frac{4\pi}{\beta}\tilde{\tau}-4\epsilon')\int_{\frac{\pi}{2}-\frac{2\pi}{\beta}\tilde{\tau}}^{\frac{\pi}{2}+\frac{2\pi}{\beta}\tilde{\tau}}du
\;[\frac{3(2u-\pi)}{{\sin}^5(2u)}-\frac{2(2u-\pi)}{{\sin}^3(2u)}
-\frac{3{\cos}(2u)}{{\sin}^4(2u)}]\Big\}
\end{align}
where $\epsilon'=\frac{\pi}{\beta}\epsilon$. Finally, we find
\begin{align}
G_{\lambda\lambda}=&\frac{V_{\mathbb R}}{2}\mathcal C\Big\{\frac{\pi}{128\epsilon^3}+(\frac{\pi}{\beta})^2\frac{\pi}{8\epsilon}+ \frac{1}{48}(\frac{\pi}{\beta})^3\tan(a)\Big(-24a-28\cot(a)+6i \pi^2a\,\cot(a)-24a\,\cot^2(a)\cr
&+6\cot(a) \,\csc^2(a)-6a\,\cot^2(a)\,\csc^2(a)-6\csc(a)\,\sec(a)-6a\,\sec^2(a)+42\cot(a)\,\zeta(3) \cr
&+48i\cot(a)(a\,\chi_2(e^{-2i a})-i \chi_3(e^{-2i a}))\Big)\Big\}
\end{align}
where $a=\frac{2\pi}{\beta}\tilde{\tau}$, $\zeta(x)$ is the Riemann $\zeta$-function, and $\chi_s$ is the Legendre $\chi$-function, which is defined as
\be
\chi_s(z)=\frac{1}{2}[{\rm Li}_s(z)-{\rm Li}_s(-z)]
\ee
Here ${\rm Li}_s(z)$ is the polylogarithm function defined by \cite{1981LL}
\be
{\rm Li}_s(z)=\sum_{k=1}^{\infty}\frac{z^k}{k^s}
\ee

$\bullet\quad \Delta=5$
\begin{align}
G_{\lambda\lambda}=&\frac{V_{\mathbb R}}{2}\mathcal C(\frac{\pi}{\beta})^7\int_{-\frac{\pi}{\beta}\tilde{\tau}+\epsilon' }^{\frac{\pi}{2}+\frac{\pi}{\beta}\tilde{\tau}-\epsilon' }d\tau_1'\int_{\frac{\pi}{2}+\frac{\pi}{\beta}\tilde{\tau}+\epsilon'}^{\pi-\frac{\pi}{\beta}\tilde{\tau}-\epsilon' }d\tau_2'\int_{-\infty}^{+\infty}dx_1'\frac{1}{(\sinh^2(x_1'-x_2')+\sin^2(\tau_1'-\tau_2'))^5}\cr
=&\frac{V_{\mathbb R}}{2}\mathcal C\Big\{ \frac{5 \pi }{65536 \epsilon ^7}+(\frac{\pi}{\beta})^2\frac{5 \pi }{6144 \epsilon ^5}+(\frac{\pi}{\beta})^4\frac{59 \pi }{9216 \epsilon ^3}+(\frac{\pi}{\beta})^6\frac{5 \pi }{64 \epsilon }\cr
&+(\frac{\pi}{\beta})^7\frac{1}{688128}\Big(387072 (i a\chi_2(e^{-2 i a})+\chi_3(e^{-2 i a}))+768 (63 i \pi ^2 a+441 \zeta (3)-358)\cr
&+\frac{7}{4} \csc ^7(a) \sec ^7(a) (-6102 a+2215 \sin (4 a)-188 \sin (8 a)+27 \sin (12 a)\cr
&+3 a (-473 \cos (4 a)-62 \cos (8 a)+9 \cos (12 a)))\Big)\Big\}
\end{align}

$\bullet\quad \Delta=4$
\begin{align}
G_{\lambda\lambda}=&\frac{V_{\mathbb R}}{2}\mathcal C(\frac{\pi}{\beta})^5\int_{-\frac{\pi}{\beta}\tilde{\tau}+\epsilon' }^{\frac{\pi}{2}+\frac{\pi}{\beta}\tilde{\tau}-\epsilon' }d\tau_1'\int_{\frac{\pi}{2}+\frac{\pi}{\beta}\tilde{\tau}+\epsilon'}^{\pi-\frac{\pi}{\beta}\tilde{\tau}-\epsilon' }d\tau_2'\int_{-\infty}^{+\infty}dx_1'\frac{1}{(\sinh^2(x_1'-x_2')+\sin^2(\tau_1'-\tau_2'))^4}\cr
=&\frac{V_{\mathbb R}}{2}\mathcal C\Big\{\frac{\pi}{1536 \epsilon^5} + (\frac{\pi}{\beta})^2\frac{\pi}{144 \epsilon^3} + (\frac{\pi}{\beta})^4\frac{\pi}{9\epsilon}+(\frac{\pi}{\beta})^5\frac{1}{1080} \csc ^5(2 a) \Big(-910 \sin (2 a)\cr
&+95 \sin (6 a)-19 \sin (10 a)+60 a (28 \cos (2 a)-5 \cos (6 a)+\cos (10 a))\Big)\Big\} \label{mimDelta4}
\end{align}

$\bullet\quad \Delta=6$
\begin{align}
G_{\lambda\lambda}=&\frac{V_{\mathbb R}}{2}\mathcal C(\frac{\pi}{\beta})^9\int_{-\frac{\pi}{\beta}\tilde{\tau}+\epsilon' }^{\frac{\pi}{2}+\frac{\pi}{\beta}\tilde{\tau}-\epsilon' }d\tau_1'\int_{\frac{\pi}{2}+\frac{\pi}{\beta}\tilde{\tau}+\epsilon'}^{\pi-\frac{\pi}{\beta}\tilde{\tau}-\epsilon' }d\tau_2'\int_{-\infty}^{+\infty}dx_1'\frac{1}{(\sinh^2(x_1'-x_2')+\sin^2(\tau_1'-\tau_2'))^6}\cr
=&\frac{V_{\mathbb R}}{2}\mathcal C\Big\{\frac{7\pi }{655360\epsilon ^9}+(\frac{\pi}{\beta})^2\frac{\pi }{8192\epsilon^7}+(\frac{\pi}{\beta})^4\frac{11 \pi}{12800\epsilon^5}+(\frac{\pi}{\beta})^6\frac{\pi}{180\epsilon^3}+(\frac{\pi}{\beta})^8\frac{16\pi}{225\epsilon}\cr
&+(\frac{\pi}{\beta})^9\frac{1}{378000}\csc^9(2a)\Big(-593334\sin(2a)-146244\sin(6a)\cr
&+840a(2376\cos(2a)+116\cos(6a)+36\cos(10a)-9\cos(14a)+\cos(18a))\cr
&-209(36\sin(10a)-9\sin(14a)+\sin(18a))\Big)\Big\}
\end{align}

The form of the mixed-state information metric for the scalar primary operator of conformal dimension $\Delta=4$ is simpler than other ones. In the next section, we attempt to find the mixed-state information metric for the scalar primary operator of conformal dimension $\Delta=4$ in $AdS_3$ gravity and compare it with the formula \eqref{mimDelta4}.

\section{Mixed-state information metric in AdS$_3$}
\label{QIMgr}
In this section we compute the mixed-state information metric for the scalar primary operator of conformal dimension $\Delta=4$ holographically.
Because of the irrelevant deformation, we will consider a massive field probing a fixed bulk geometry. The gravity dual to a thermal state in $CFT_2$
is the BTZ black string,
\be
ds^2=-(\frac{r^2-r_+^2}{l^2})d t^2+(\frac{l^2}{r^2-r_+^2})d r^2+r^2 d\phi^2 \label{btzm}
\ee
where $l$ is the AdS radius and $r_+$ is the horizon radius. The range of the coordinates are $-\infty<t, \phi<\infty$, $r_+<r<\infty$,
the Hawking temperature is ${1\over\beta}=\frac{r_+}{2\pi l^2}$, demanding the Euclidean BTZ black string is smooth at the horizon we make the identification $\tau\sim\tau+\beta$. For simplicity, we will set $l=r_+=1$, the metric \eqref{btzm} can be transformed to
\be
ds^2=\frac{dZ^2}{(1-Z^2)Z^2}+\frac{1-Z^2}{Z^2}d\tau^2+\frac{d\phi^2}{Z^2} \label{btz}
\ee
The metric \eqref{btz} has appeared in \cite{Trivella:2016brw}, this means that our holographic computation can be based on the method used in
\cite{Trivella:2016brw}.

It is clear that the fidelity can be interpreted as a combination of partition functions
\be
F(\rho_1,\rho_2)=\frac{Z_2}{\sqrt{Z_0Z_1}}\label{fi1}
\ee
where $Z_0$ is the partition function of a pure CFT, $Z_1$ is the partition function of the deformed CFT and $Z_2$ is the partition function of the pure
CFT deformed only on the interval $[\pi+\tilde\tau,2\pi-\tilde\tau]$. In the semi-classical gravity limit, we have $Z_k={\exp}(-I_k)$ where $I_k$ is the
on-shell action of the bulk geometry dual to the corresponding field theory configuration.

To obtain the information metric $G_{\lambda\lambda}$ in \eqref{def}, we need performing a perturbative expansion in infinitesimal $\delta \lambda$.
The detailed analysis has been done in \cite{Trivella:2016brw}, which finds
\be
-\log Z_k=I_{AdS}+\delta I_k+\O(\delta \lambda^4)
\ee
where $I_{AdS}$ is the on shell action of pure Einstein theory with the cosmological constant $\Lambda<0$ and $\delta I_k$ is the on shell action of
the scalar fields probing a fixed AdS background. Specifically,
\be
\delta I_k=\eta\int_{\partial \mathcal M_\epsilon}\sqrt{\gamma_0}n_\mu g^{\mu\nu}\Phi_k\partial_\nu\Phi_k \label{osac}
\ee
where $\mathcal M_\epsilon$ is the regularized AdS, $n_\mu$ is the unit normal vector at the cut-off surface $\partial \mathcal M_\epsilon$ and
$\gamma_0$ is the determinant of the induced metric on $\partial \mathcal M_\epsilon$.

We can write the fidelity as
\be
F=\exp(-\delta I_2+{1\over2}\delta I_1)\label{fi2}
\ee
Next, we need to find the scalar field $\Phi_k$ obeying the equation of motion, it can be obtained using the bulk to boundary propagator.  Firstly
we note that the scalar field dual to an primary operator $\O$ in CFT$_2$ satisfies the following boundary condition
\be
\lim_{Z\rightarrow 0}Z^{\Delta-2}\Phi_k(Z,\tau,\phi)\equiv\tilde\Phi_k(\tau,\phi)=\delta\lambda\,s_k(\tau)
\ee
where
\begin{align}
s_0(\tau)&=0\cr
s_1(\tau)&=1\cr
s_2(\tau)&=\begin{cases}
0\;\;\;\;-\tilde\tau\leq\tau\leq\pi+\tilde\tau\\
1\;\;\;\;\pi+\tilde\tau\leq\tau\leq2\pi-\tilde\tau
\end{cases}
\end{align}
Secondly, the metric \eqref{btz} can be written as pure AdS under the following coordinate transformation
\begin{align}
x&=\sqrt{1-Z^2}\cos\tau e^\phi\cr
y&=\sqrt{1-Z^2}\sin\tau e^\phi\cr
z&=Z e^\phi \label{ctransf}
\end{align}
The pure AdS in Poincar\'e coordinates is
\be
ds^2=\frac{dx^2+dy^2+dz^2}{z^2}
\ee
Then the scalar field $\Phi_k$ is constructed by the following integral
\be
\Phi_k(z,x,y)=\delta\lambda\,c_\Delta\int dx'dy'\Big[\frac{z}{z^2+(x-x')^2+(y-y')^2}\Big]^\Delta s_k(x',y')\,(x'^2+y'^2)^{\Delta-2\over 2}
\ee
with the normalization constant $c_\Delta=\frac{\Delta-1}{\pi}$, where we used the fact that the source function $s(x)$ is generally not invariant under the conformal transformation. So, the scalar field $\Phi_1$ dual to the primary operator $\O$ with conformal dimension $\Delta=4$ is
\be
\Phi_1(z,x,y)=\delta\lambda\frac{2(x^2+y^2)+z^2}{2z^2}
\ee
The scalar field $\Phi_2$ is
\begin{align}
\Phi_2(z,x,y)&=\delta\lambda\,c_4\int_{-\infty}^0dy'\int_{y'/\tilde t}^{-y'/\tilde t}dx'\frac{z^4}{(z^2+(x-x')^2+(y-y')^2)^4}(x'^2+y'^2)\cr
&=\delta\lambda\,c_4\int_0^{+\infty}dy'\int_{-y'/\tilde t}^{y'/\tilde t}dx'\frac{z^4}{(z^2+(x-x')^2+(y+y')^2)^4}(x'^2+y'^2) \label{phi2}
\end{align}
The region of the integral \eqref{phi2} can be understood by the Fig.6 in \cite{Trivella:2016brw}, and $\tilde t=\tan \tau$. With some efforts, we find that the final form of the scalar field $\Phi_2$ is
\be
\Phi_2(z,x,y)=\frac{\delta \lambda}{\pi}[h(x,y,z)+h(-x,y,z)+\frac{2(x^2+y^2)+z^2}{2z^2}(\frac{\pi}{2}-\tilde\tau)]
\ee
where
\begin{align}
h(x,y,z)&=\frac{1}{16}\Big(\frac{3(\tilde t x+y)^3(x-\tilde t y)}{\big((y+\tilde t x)^2+(1+{\tilde t}^2)z^2\big)^2}+
\frac{7(y+\tilde t x)(-x+\tilde t y)}{(y+\tilde t x)^2+(1+{\tilde t}^2)z^2}\Big)\cr
&+(y+\tilde t x)\Big(8(x^2+y^2)(y+\tilde t x)^4+4(y+\tilde t x)^2\big(2x y \tilde t+y^2(6+5\tilde t^2)+x^2(5+6{\tilde t}^2)\big)z^2\cr
&+5(1+{\tilde t}^2)\big(4xy\tilde t+y^2(5+3\tilde t^2)+x^2(3+5\tilde t^2)\big)z^4+9(1+\tilde t^2)^2z^6\Big)\cr
&\frac{\big(-\pi+2\arctan \frac{\tilde t y-x}{\sqrt{(y+\tilde t x)^2+(1+\tilde t^2)z^2}}\big)}{32z^2\big((y+\tilde t x)^2+(1+\tilde t^2)z^2\big)^{5/2}}
\end{align}
To compute the on-shell action \eqref{osac}£¬ we need putting a cut off at $z=\epsilon$ as a regularization for AdS space, we then have
\be
\delta I_k=-\frac{1}{2\kappa^2}\int dxdy \frac{1}{\epsilon}\Phi_k\partial_z\Phi_k
\ee
According to the coordinate transformation \eqref{ctransf}, the above integral become to
\be
\delta I_k=-\frac{1}{2\kappa^2}\int d\phi d\tau e^{2\phi}\lim_{\epsilon\to0}\frac{1}{\epsilon}\Phi_k\partial_z\Phi_k\vert_{z=\epsilon}
\ee
For the action $\delta I_1$, we note that
\be
\frac{1}{\epsilon}\Phi_1\partial_z\Phi_1=-(\delta\lambda)^2\big[\frac{2e^{4\phi}}{\epsilon^6}-\frac{3e^{2\phi}}{\epsilon^4}+\frac{1}{\epsilon^2}\big]\label{I1}
\ee
So the integrand in $\delta I_1$ is divergent under the limit $\epsilon\to0$. By using the counterterms, we simply subtract the power divergences in
\eqref{I1} \cite{Trivella:2016brw}. With this argument, the contribution of the bulk configuration $\Phi_1$ to the information metric is absent in our case.

Next, we calculate the on-shell action $\delta I_2$, the first piece is
\begin{align}
\mathcal F(\tau)&=\lim_{\epsilon\to0}\frac{1}{\epsilon}\Phi_2\partial_z\Phi_2=-\frac{(\delta \lambda)^2}{\pi^2}\frac{e^{-2\phi}}{49152}\csc^6(\tau-\tilde\tau)\csc ^6(\tau +\tilde\tau)\cr
&\left(\arctan(\cot(\tau -\tilde \tau))-\arctan(\cot(\tau +\tilde\tau ))-2\tilde\tau+2\pi\right)\cr
&\Big(-6\cos(8\tau)(19\sin(4\tilde\tau)+6\pi\cos(4\tilde\tau))\cr
&-4\cos(6\tau)(-171\sin(2\tilde\tau)+5\sin(6\tilde\tau)+42\pi\cos(6\tilde\tau))\cr
&+12\cos(4\tau)(10\sin(4\tilde\tau)+12\sin(8\tilde\tau)+84\pi\cos(4\tilde\tau)-3\pi\cos(8\tilde\tau)-45\pi)\cr
&+12\cos(2\tau)(6\pi\cos(2\tilde\tau)(6\cos(4\tau)+6\cos(4\tilde\tau)-31)-215\sin(2\tilde\tau)-72\sin(6\tilde\tau)-\sin(10\tilde\tau))\cr
&-768\sin^6(\tau-\tilde\tau)(3\cos(2(\tau +\tilde\tau))+7)\arctan(\cot(\tau+\tilde\tau))\cr
&+768\sin^6(\tau +\tilde\tau)(3\cos(2\tau-2\tilde\tau)+7)\arctan(\cot(\tau-\tilde\tau))\cr
&+1935\sin(4\tilde\tau)+36\sin(8\tilde\tau)+\sin(12\tilde\tau)-540\pi\cos(4\tilde\tau)+1680\pi\Big) \label{I2}
\end{align}
Because the $\tau$ integral ranges over $\tau\in[-\pi+\tilde\tau,-\tilde\tau]$ and the integrand \eqref{I2} is even, we restrict to $\tau\in[-\pi/2,-\tilde\tau]$. Due to the divergence arising from the interface at $\tau=-\tilde \tau$ and $\tau=-\pi+\tilde \tau$, we shift the variable
$\tau'=\tau-\epsilon$. We find that
\begin{align}
\delta I_2=&-\frac{(\delta \lambda)^2 V_{\mathbb R}}{\pi^2 \kappa^2}\int_{-\pi/2-\epsilon}^{-\tilde\tau-\epsilon}d\tau \mathcal F(\tau)\cr
=&\frac{(\delta \lambda)^2 V_{\mathbb R}}{\pi^2 \kappa^2}\Big[\frac{\pi^2}{16\epsilon^5}+\frac{\pi^2}{24\epsilon^3}+\frac{\pi^2}{24\epsilon}\cr
&+\frac{\pi}{11520}\csc^5(2\tilde\tau)\Big(-910\sin(2\tilde\tau)-19\sin(10\tilde\tau)+95\sin(6\tilde\tau)\cr
&+60\tilde\tau(28\cos(2\tilde\tau)-5\cos(6\tilde\tau)+\cos(10\tilde\tau))\Big)\Big] \label{I2f}
\end{align}
Putting the ingredients \eqref{def}, \eqref{fi2}, \eqref{I1} and \eqref{I2f} together, we read off the mixed-state information metric for the primary operator with conformal dimension $\Delta=4$ as
\begin{align}
G_{\lambda\lambda}=&\frac{V_{\mathbb R}}{16\kappa^2\epsilon^5}+\frac{V_{\mathbb R}}{24\kappa^2\epsilon^3}+\frac{V_{\mathbb R}}{24\kappa^2\epsilon}\cr
&+\frac{V_{\mathbb R}}{\pi\kappa^2}\frac{1}{11520}\csc^5(2\tilde\tau)\Big(-910\sin(2\tilde\tau)-19\sin(10\tilde\tau)+95\sin(6\tilde\tau)\cr
&+60\tilde\tau(28\cos(2\tilde\tau)-5\cos(6\tilde\tau)+\cos(10\tilde\tau))\Big)\label{mimDelta4g}
\end{align}
We can see that there is a precise matching between the CFT$_2$ result \eqref{mimDelta4} and the result \eqref{mimDelta4g} in the holographic set-up.

\section{Conclusion and Discussion}
\label{dis}
In this note we compute the mixed-state information metric for a thermal CFT on $\mathbb R^{1,1}$ deformed by the scalar primary operators of conformal
dimension $\Delta=3, 4, 5, 6$. To make the problem tractable, we consider the Hamiltonians of the two mixed states are commutative and the temperatures
of both states are the same. With these assumptions, we find that the problem is actually similar to the evaluation of the pure-state information metric in CFT. Interestingly we find that the mixed-state information metrics for the primary operators with conformal dimension $\Delta=3,5$ contain the polylogarithm functions ${\rm Li}_2(z)$ and ${\rm Li}_3(z)$. While the mixed-state information metrics for the primary operators with conformal dimension $\Delta=4,6$ are just composed of elementary functions. Making use of the method developed in \cite{Trivella:2016brw}, we also compute the mixed-state information metric for the primary operator with conformal dimension $\Delta=4$ in the bulk AdS$_3$. We find that there is an exact duality between the results in CFT$_2$ and AdS$_3$.

Despite our example supporting the $AdS_3/CFT_2$ correspondence is established in the particular temperature $\beta=2\pi$, it is straightforward to verify the $AdS_3/CFT_2$ correspondence for an arbitrary temperature. Moreover, it would be interesting to check the $AdS_3/CFT_2$ correspondence by the mixed-state information metric for the primary operators with conformal dimension $\Delta=3,5$. From the study of \cite{Bak:2017rpp}, we think that the mixed-state information metric for a thermal CFT on $\mathbb R\times \rm S$ can be reproduced by the holographic theory in BTZ black hole \cite{Banados:1992wn}. Furthermore, the method of \cite{Trivella:2016brw} should allow us to compute the quantum information metric for a 2d CFT deformed
by the fermionic primary operators holographically.

Another possible extension is to investigate the mixed-state information metric for a thermal CFT on $\mathbb R^{1,d}$, however the main difficulty is that the two point function of the thermal CFT on $\mathbb R^{1,d}$ is not available \cite{Iliesiu:2018fao}. So we need more feasible ideas to solve this problem. Besides the higher dimensional generalization, it is also natural to study the quantum information metric in 1d CFT, for example conformal quantum mechanics \cite{deAlfaro:1976vlx}, of which the two-point function has been found in \cite{Chamon:2011xk}. More importantly, it would be nice to understand the exotic $AdS_2/CFT_1$ correspondence \cite{Almheiri:2014cka} by quantum information metric.

\section*{Acknowledgements}
The research of X. B. Xu is supported by NSFC grants No$.$ 11747017. The work of S. Q. Lan is supported by Department of Education of Guangdong Province, China (Grant Nos$.$ 2017KQNCX124). The work of G. Q. Li is supported by Natural Science Foundation of Guangdong Province, China, under Grant No$.$ 2016A030307051 and Department of Education of Guangdong Province, China (Grant Nos$.$ 2017KZDXM056). The work of J. X. Mo is supported by NSFC grants No$.$ 11605082 and Natural Science Foundation of Guangdong Province, China, under Grant No$.$ 2016A030310363.

\bibliographystyle{JHEP}
\bibliography{j}

\providecommand{\href}[2]{#2}\begingroup\raggedright\begin{thebibliography}{10}

\bibitem{2010qcqi}
M.~A. Nielsen and I.~L. Chuang, {\em {Quantum Computation and Quantum
  Information}}.
\newblock Cambridge University Press, 2010.

\bibitem{Maldacena:1997re}
J.~M. Maldacena, {\it {The Large N limit of superconformal field theories and
  supergravity}},  {\em Int. J. Theor. Phys.} {\bf 38} (1999) 1113--1133,
  [\href{http://arxiv.org/abs/hep-th/9711200}{{\tt hep-th/9711200}}]. [Adv.
  Theor. Math. Phys.2,231(1998)].

\bibitem{Gubser:1998bc}
S.~S. Gubser, I.~R. Klebanov, and A.~M. Polyakov, {\it {Gauge theory
  correlators from noncritical string theory}},  {\em Phys. Lett.} {\bf B428}
  (1998) 105--114, [\href{http://arxiv.org/abs/hep-th/9802109}{{\tt
  hep-th/9802109}}].

\bibitem{Witten:1998qj}
E.~Witten, {\it {Anti-de Sitter space and holography}},  {\em Adv. Theor. Math.
  Phys.} {\bf 2} (1998) 253--291,
  [\href{http://arxiv.org/abs/hep-th/9802150}{{\tt hep-th/9802150}}].

\bibitem{Ryu:2006bv}
S.~Ryu and T.~Takayanagi, {\it {Holographic derivation of entanglement entropy
  from AdS/CFT}},  {\em Phys. Rev. Lett.} {\bf 96} (2006) 181602,
  [\href{http://arxiv.org/abs/hep-th/0603001}{{\tt hep-th/0603001}}].

\bibitem{Hartman:2013qma}
T.~Hartman and J.~Maldacena, {\it {Time Evolution of Entanglement Entropy from
  Black Hole Interiors}},  {\em JHEP} {\bf 05} (2013) 014,
  [\href{http://arxiv.org/abs/1303.1080}{{\tt arXiv:1303.1080}}].

\bibitem{Susskind:2014rva}
L.~Susskind, {\it {Computational Complexity and Black Hole Horizons}},  {\em
  Fortsch. Phys.} {\bf 64} (2016) 44--48,
  [\href{http://arxiv.org/abs/1403.5695}{{\tt arXiv:1403.5695}}].

\bibitem{MIyaji:2015mia}
M.~Miyaji, T.~Numasawa, N.~Shiba, T.~Takayanagi, and K.~Watanabe, {\it
  {Distance between Quantum States and Gauge-Gravity Duality}},  {\em Phys.
  Rev. Lett.} {\bf 115} (2015), no.~26 261602,
  [\href{http://arxiv.org/abs/1507.07555}{{\tt arXiv:1507.07555}}].

\bibitem{Stanford:2014jda}
D.~Stanford and L.~Susskind, {\it {Complexity and Shock Wave Geometries}},
  {\em Phys. Rev.} {\bf D90} (2014), no.~12 126007,
  [\href{http://arxiv.org/abs/1406.2678}{{\tt arXiv:1406.2678}}].

\bibitem{Harlow:2018fse}
D.~Harlow, {\it {TASI Lectures on the Emergence of the Bulk in AdS/CFT}},
  \href{http://arxiv.org/abs/1802.01040}{{\tt arXiv:1802.01040}}.

\bibitem{Caputa:2017yrh}
P.~Caputa, N.~Kundu, M.~Miyaji, T.~Takayanagi, and K.~Watanabe, {\it {Liouville
  Action as Path-Integral Complexity: From Continuous Tensor Networks to
  AdS/CFT}},  {\em JHEP} {\bf 11} (2017) 097,
  [\href{http://arxiv.org/abs/1706.07056}{{\tt arXiv:1706.07056}}].

\bibitem{Braunstein:1994zz}
S.~L. Braunstein and C.~M. Caves, {\it {Statistical distance and the geometry
  of quantum states}},  {\em Phys. Rev. Lett.} {\bf 72} (1994) 3439--3443.

\bibitem{Spehner:2014jmp}
D.~Spehner, {\it {Quantum correlations and distinguishability of quantum
  states}},  {\em Journal of Mathematical Physics} {\bf 55} (2014), no.~7
  075211, [\href{http://arxiv.org/abs/1407.3739}{{\tt arXiv:1407.3739}}].

\bibitem{Bak:2003jk}
D.~Bak, M.~Gutperle, and S.~Hirano, {\it {A Dilatonic deformation of AdS(5) and
  its field theory dual}},  {\em JHEP} {\bf 05} (2003) 072,
  [\href{http://arxiv.org/abs/hep-th/0304129}{{\tt hep-th/0304129}}].

\bibitem{Bak:2015jxd}
D.~Bak, {\it {Information metric and Euclidean Janus correspondence}},  {\em
  Phys. Lett.} {\bf B756} (2016) 200--204,
  [\href{http://arxiv.org/abs/1512.04735}{{\tt arXiv:1512.04735}}].

\bibitem{Trivella:2016brw}
A.~Trivella, {\it {Holographic Computations of the Quantum Information
  Metric}},  {\em Class. Quant. Grav.} {\bf 34} (2017), no.~10 105003,
  [\href{http://arxiv.org/abs/1607.06519}{{\tt arXiv:1607.06519}}].

\bibitem{Bak:2017rpp}
D.~Bak and A.~Trivella, {\it {Quantum Information Metric on $\mathbb{R} \times
  S^{d-1}$}},  {\em JHEP} {\bf 09} (2017) 086,
  [\href{http://arxiv.org/abs/1707.05366}{{\tt arXiv:1707.05366}}].

\bibitem{Chen:2018vkw}
C.-B. Chen, W.-C. Gan, F.-W. Shu, and B.~Xiong, {\it {Quantum information
  metric of conical defect}},  {\em Phys. Rev.} {\bf D98} (2018), no.~4 046008,
  [\href{http://arxiv.org/abs/1804.08358}{{\tt arXiv:1804.08358}}].

\bibitem{Iliesiu:2018fao}
L.~Iliesiu, M.~Kolo\u{g}lu, R.~Mahajan, E.~Perlmutter, and D.~Simmons-Duffin,
  {\it {The Conformal Bootstrap at Finite Temperature}},
  \href{http://arxiv.org/abs/1802.10266}{{\tt arXiv:1802.10266}}.

\bibitem{1981LL}
L.~Lewin, {\em {Polylogarithms and Associated Functions}}.
\newblock North Holland, New York, 1981.

\bibitem{Banados:1992wn}
M.~Banados, C.~Teitelboim, and J.~Zanelli, {\it {The Black hole in
  three-dimensional space-time}},  {\em Phys. Rev. Lett.} {\bf 69} (1992)
  1849--1851, [\href{http://arxiv.org/abs/hep-th/9204099}{{\tt
  hep-th/9204099}}].

\bibitem{deAlfaro:1976vlx}
V.~de~Alfaro, S.~Fubini, and G.~Furlan, {\it {Conformal Invariance in Quantum
  Mechanics}},  {\em Nuovo Cim.} {\bf A34} (1976) 569.

\bibitem{Chamon:2011xk}
C.~Chamon, R.~Jackiw, S.-Y. Pi, and L.~Santos, {\it {Conformal quantum
  mechanics as the CFT$_1$ dual to AdS$_2$}},  {\em Phys. Lett.} {\bf B701}
  (2011) 503--507, [\href{http://arxiv.org/abs/1106.0726}{{\tt
  arXiv:1106.0726}}].

\bibitem{Almheiri:2014cka}
A.~Almheiri and J.~Polchinski, {\it {Models of AdS$_{2}$ backreaction and
  holography}},  {\em JHEP} {\bf 11} (2015) 014,
  [\href{http://arxiv.org/abs/1402.6334}{{\tt arXiv:1402.6334}}].

\end{thebibliography}\endgroup
\end{document}